April 2004

# Indicators of the Energy Error in the Linac to Booster Transfer

Xi Yang and James MacLachlan

*Fermi National Accelerator Laboratory*

Box 500, Batavia IL 60510

## Abstract

The match between the Linac beam energy and the energy determined by the bending field of the Booster magnets is crucial for rf capture, beam quality, and the transmission efficiency in a Booster cycle. The observation of the injection energy match is important for injection tuning. Several signals, such as phase shift drive (PSD), radial position error (RPOS), synchrotron phase (SPD), and fast phase error (FPERR), provide consistent information on the energy match and can be used for a injection match tuning.

## Introduction

The Booster accelerates protons ($H^+$) from 400 MeV to 8 GeV after the $H^-$ beam injected from the Linac is stripped to protons. The rf frequency changes from 37.9 MHz to 52.9 MHz to accelerate the beam in a Booster cycle; simultaneously the magnetic field changes to keep the beam on the central orbit. During the injection period it is important to match the bending field of the Booster magnets with the injected beam energy; in addition, the rf frequency must equal the product of the circulation frequency and the Booster harmonic number for the beam to be accelerated properly.

## Observations

Observations of the injection process are reported for two different energy match conditions, a mismatch and a good match. Each set includes two different beam intensities. In the mismatched condition, for the extracted beam intensity of $4.7 \times 10^{12}$ protons, there was a spike in PSDRV, indicating that the phase feedback was initiated about 30 μs after injection; also there was a broader spike in the SPD signal after initiation of feedback, as shown in Fig. 1(a). RPOS indicates that the beam was injected on an orbit several millimeters outside of the desired orbit, as shown in Fig. 1(b). When PSDRV is positive before transition, the synchrotron phase is reduced and the beam gets less accelerating voltage. The spike with a millisecond width and three-volt amplitude in PSDRV shown in Fig. 1(c) indicates that the injected beam energy was too high. This is consistent with the observation made from RPOS; it took about 400 μs after injection for the beam to move inward onto the right orbit, as shown in Fig. 1(d). A spike with a width of 10-20 μs and a peak of 4.6 volts in FPERR was also observed right after injection, as shown in Fig. 1(e), and also indicates that the injected beam energy was higher than desired. The injected beam was bunched by the rf over 20 μs after injection, initially at 0° synchronous phase. Fig. 1(g) is has the same quantities as Fig. 1(f), but only the time period surrounding the arrival of linac beam is plotted. On this expanded time scale it is apparent that the beam picks up enough 37.9 MHz modulation in the first 20 μs to yield a significant synchrotron phase signal even though the last few turns of injected beam are not yet captured.

In the matched condition, for the extracted beam intensity of $4.7 \times 10^{12}$ protons, there was a spike in PSDRV shown in Fig. 2a with a much smaller amplitude on the rising edge than that shown for the unmatched condition shown in Fig. 2(a). Also the counterpart of the broad SPD spike in Fig. 1(a) is scarcely evident. The RPOS near zero seen in Fig. 2(b) indicates that the beam was nearly on the desired orbit. The counterpart of the PSDRV spike in Fig. 1(c) also disappeared, as shown in Fig. 2(c). The orbit movement shown in Fig. 2(d) was smaller than that shown in Fig. 1(d). The FPERR spike right after injection was negative, as shown in Fig. 2(e). Both RPOS and FPERR indicate that the injected beam energy was close to the desired. In the mismatched situation, for the extracted beam intensity of $0.8 \times 10^{12}$ protons, the spike in PSDRV indicates the feedback turns on at about 65 μs after injection, as shown in Fig. 3(a). It took a longer time before the feedback was initiated than in the situation shown in Fig. 1(a). The situation in this instance is that the time taken starting feedback is inversely proportional to beam intensity. The feedback-on time is independent of the beam intensity only when the beam gate on-time (B:TFBON) is set at a time later than the beam gate time triggered by the beam intensity. RPOS also indicates that the injected beam energy was higher than the desired. Figs. 3(b)-(f) are similar to Figs 1(b)-(f). PSDRV is dependent upon the beam intensity, as can be seen by comparing Fig. 1(c) and Fig. 3(c).

The matched condition for the extracted beam intensity of $0.4\times10^{12}$ protons, shown in Figs. 4(a)-(f), is rather similar to the matched condition for intensity of $4.7 \times 10^{12}$ shown in Figs. 2(a)-(f).

## Comments

PSDRV, RPOS, the SPD signal, and FPERR can be used for the injection match tuning. Since the synchronous phase of the beam is moving to a higher value so that the beam gets more effective accelerating voltage when PSDRV goes in the negative direction at injection, the sign and amplitude of PSDRV can be used to determine that the relationship between the injected beam energy and the central orbit energy for the Booster. RPOS goes positive when the injected beam energy is higher than the desired one, as does FPERR. Also, at mismatched injection energy, the SPD signal has a spike after feedback starts, which does not appear in the matched condition.

A feature of Fig. 1(f) worth remarking for other applications of the SPD is the small sudden beam loss just after 4.3 ms reflecting the creation of a notch in the batch to accommodate kicker rise time. At this time one can see that the previously quiet SPD signal is disturbed by the presence of revolution harmonics other than the rf fundamental. For such purposes as measuring the synchrotron oscillation frequency, it would be helpful to have a pure signal to represent the fundamental rf component of the beam current. Perhaps it will be possible to improve the bandpass filter in the beam current input to the SPD sufficiently to reduce the unwanted circulation harmonics to a negligible level.

The match between the injected beam energy and the central orbit energy for the Booster is crucial for the beam capture efficiency, beam quality, transmission efficiency, etc. Several diagnostic signals have been used for the injection match tuning in Booster. These signals have been observed to be consistent in general, but reliance on any single one increases the likelihood of misinterpretation arising from some unexpected condition. The recently installed SPD is extremely sensitive to energy mismatch at injection and can be used to improve the injection match.

## Acknowledgement

Special thanks should be given to Bill Pellico for his help in understanding the operational issues related to the injection.

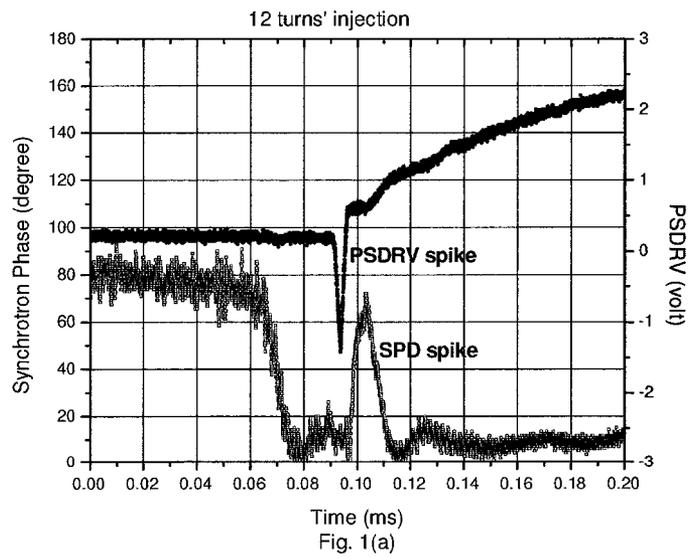

Fig. 1(a)

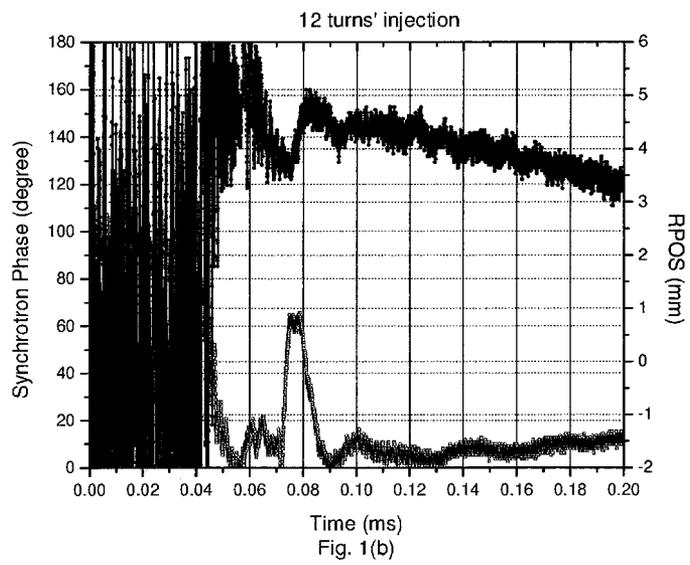

Fig. 1(b)

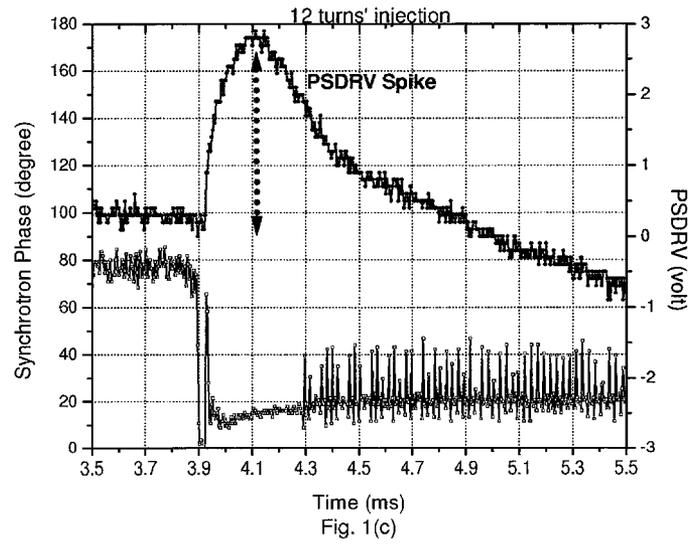

Fig. 1(c)

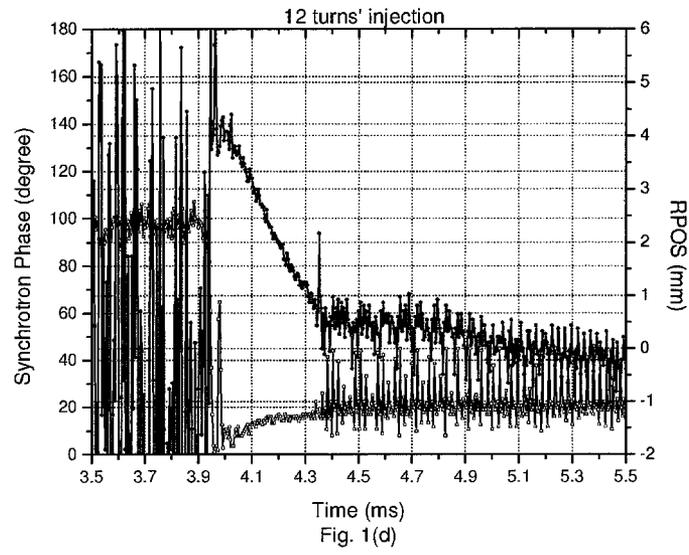

Fig. 1(d)

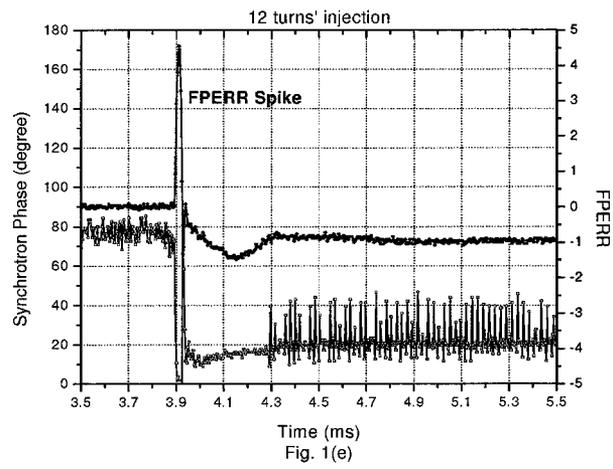

Fig. 1(e)

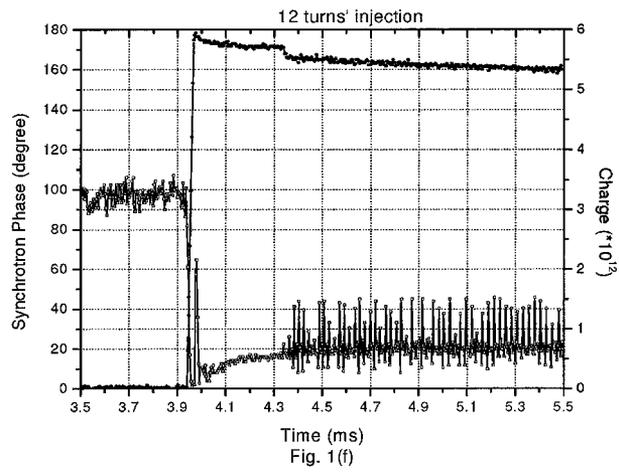

Fig. 1(f)

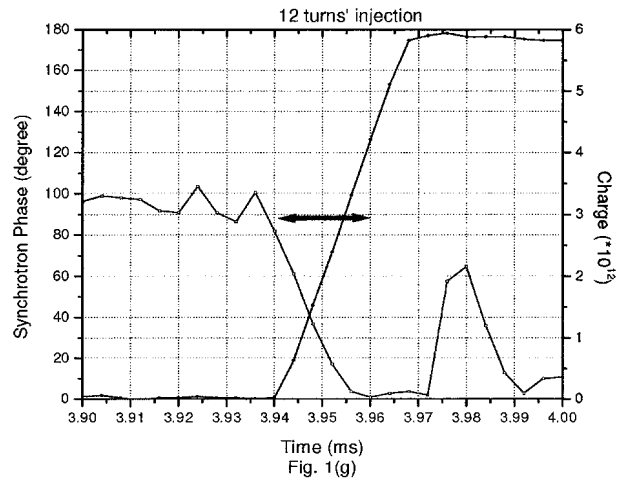

Fig. 1 The data were taken in a mismatched situation at injection for the extracted beam intensity of $4.7 \times 10^{12}$ protons.

In the first 100 μs after injection,

Fig. 1(a) the SPD signal and PSDRV *vs.* time.

Fig. 1(b) the SPD signal and RPOS *vs.* time.

In the first 1.6 ms after injection,

Fig. 1(c) the SPD signal and PSDRV *vs.* time.

Fig. 1(d) the SPD signal and RPOS *vs.* time.

Fig. 1(e) the SPD signal and FPERR *vs.* time.

Fig. 1(f) the SPD signal and charge *vs.* time.

In the first 60 μs after injection,

Fig. 1(g) the SPD signal and charge *vs.* time. The red arrow indicates the time taken for the start of synchrotron phase measurement.

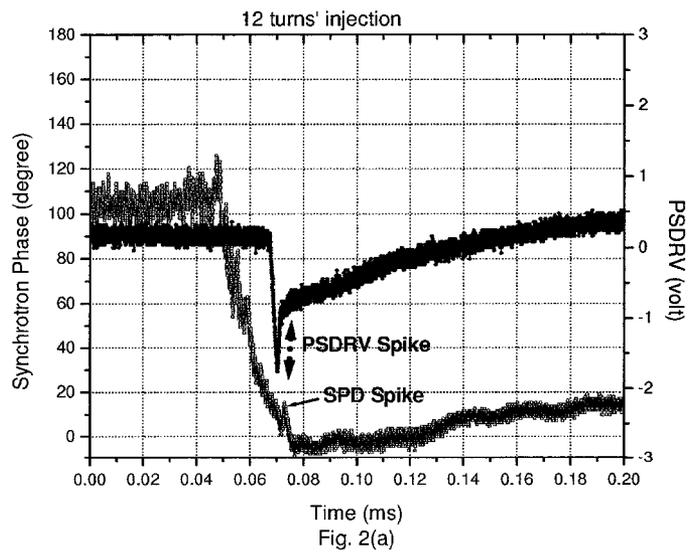

Fig. 2(a)

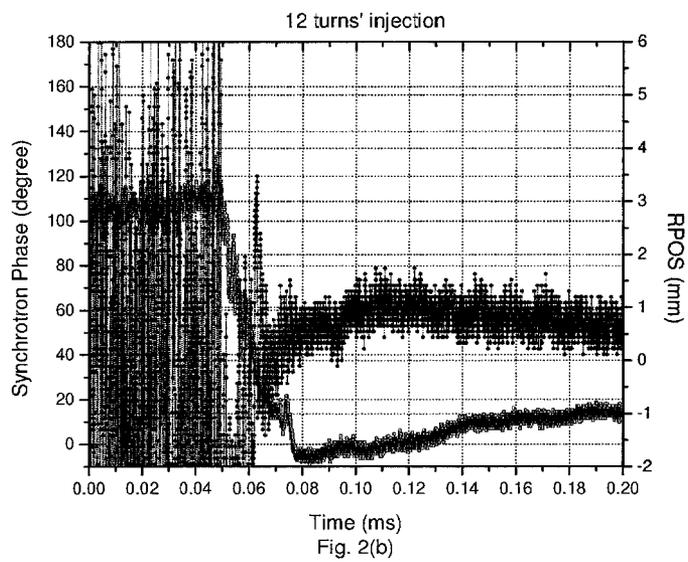

Fig. 2(b)

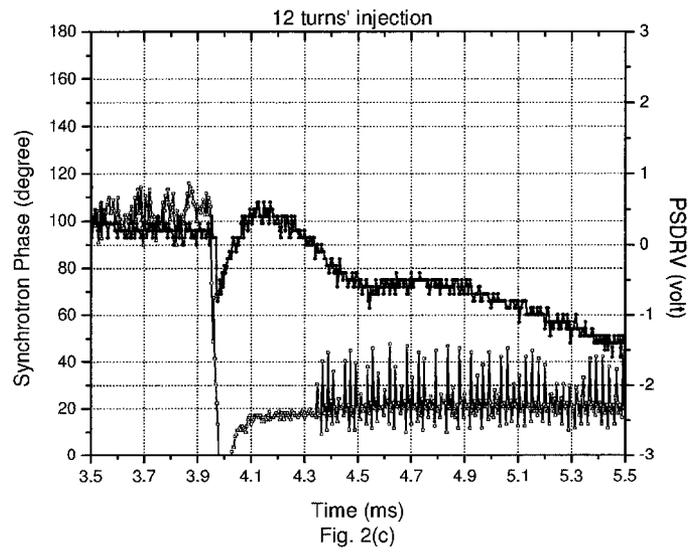

Fig. 2(c)

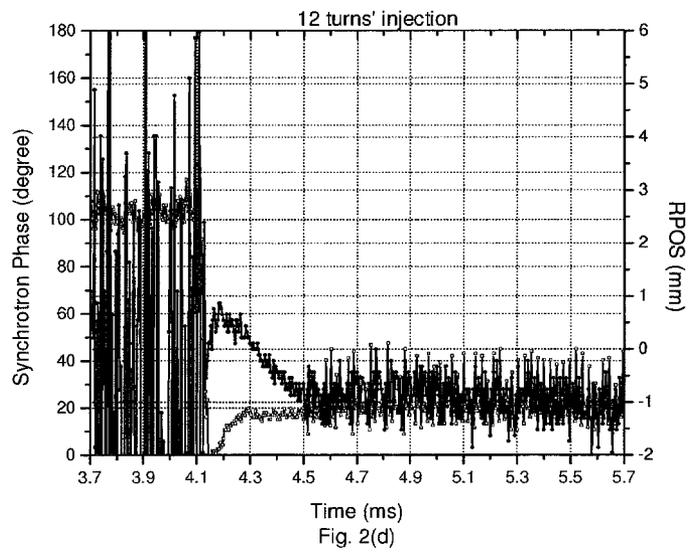

Fig. 2(d)

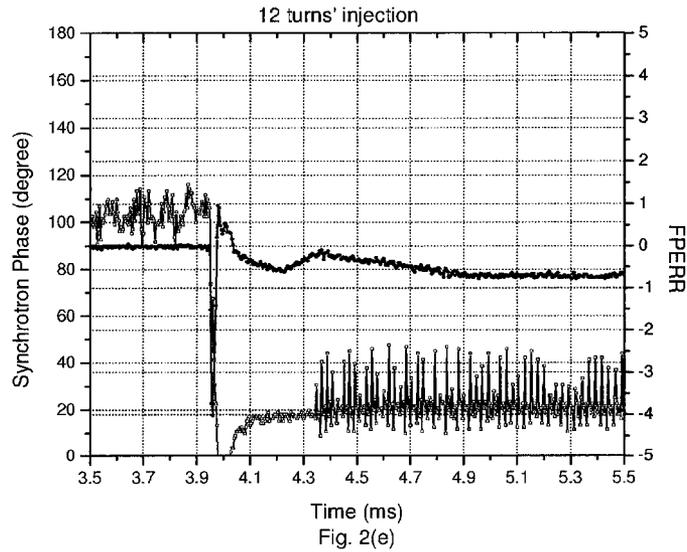

Fig. 2(e)

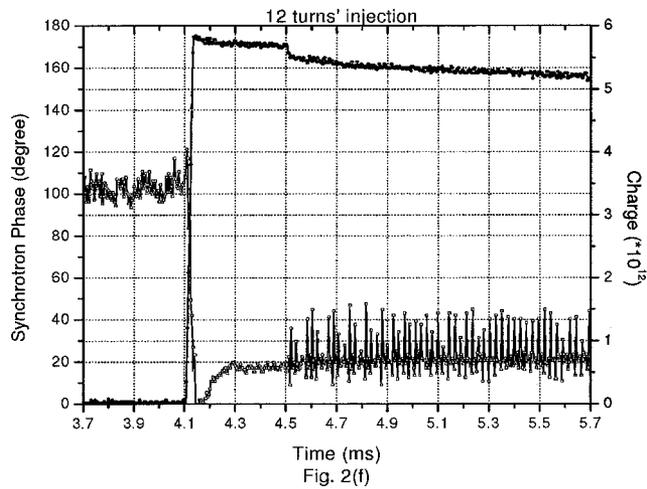

Fig. 2(f)

Fig. 2 The data were taken in the matched situation at injection for the extracted beam intensity of $4.7 \times 10^{12}$ protons.

In the first 100 μs after injection,

Fig. 2(a) the SPD signal and PSDRV *vs.* time.

Fig. 2(b) the SPD signal and RPOS *vs.* time.

In the first 1.6 ms after injection,

Fig. 2(c) the SPD signal and PSDRV *vs.* time.

Fig. 2(d) the SPD signal and RPOS *vs.* time.

Fig. 2(e) the SPD signal and FPERR *vs.* time.

Fig. 2(f) the SPD signal and charge *vs.* time.

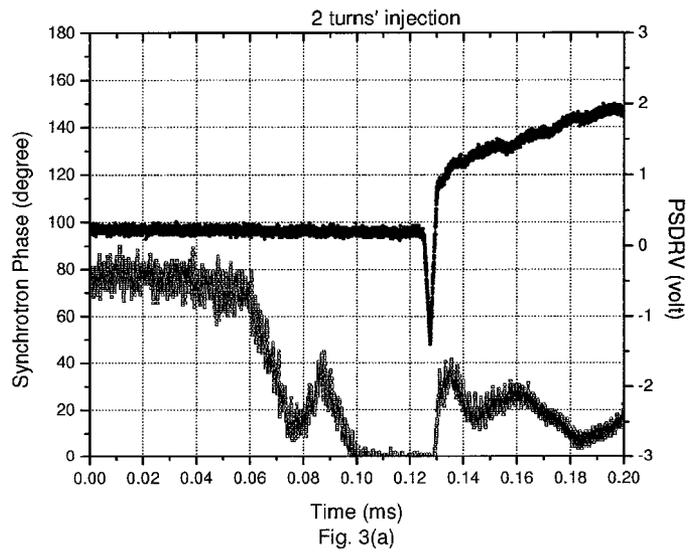

Fig. 3(a)

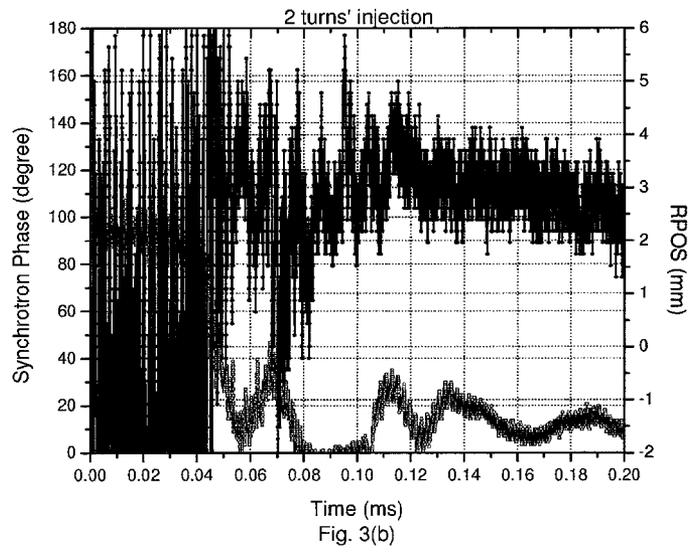

Fig. 3(b)

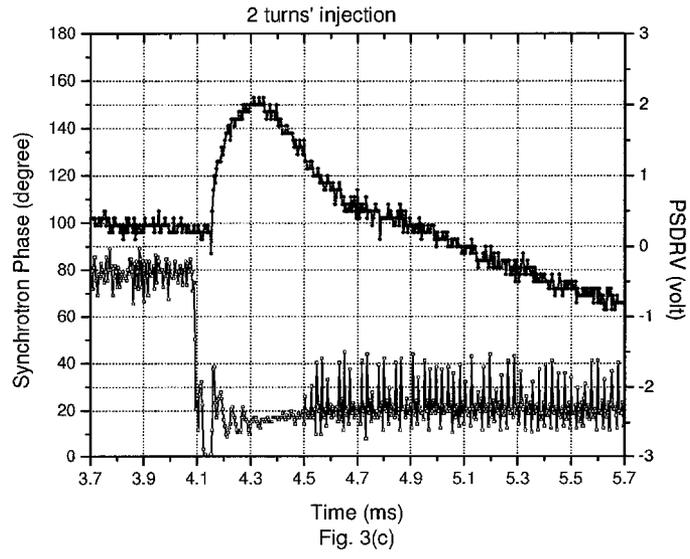

Fig. 3(c)

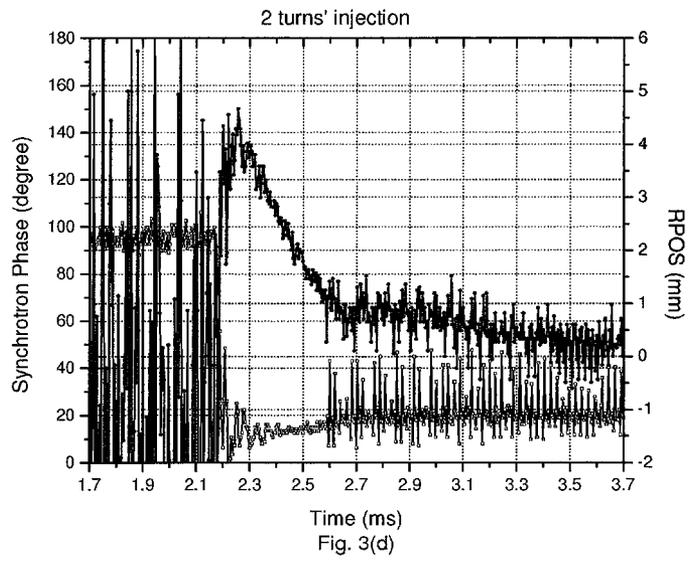

Fig. 3(d)

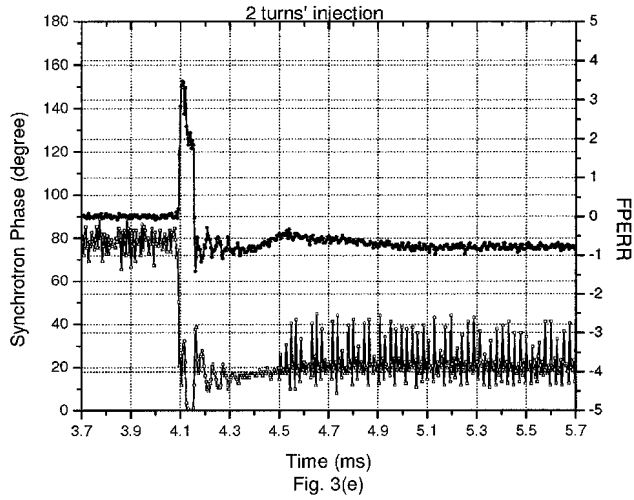

Fig. 3(e)

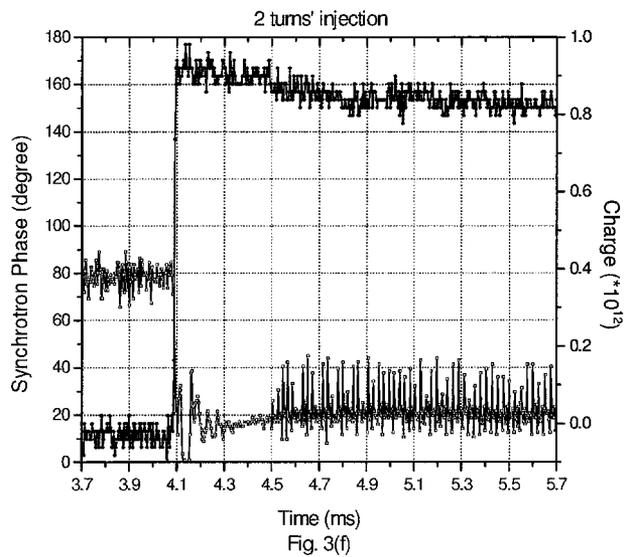

Fig. 3(f)

Fig. 3 The data were taken in the mismatched situation at injection for the extracted beam intensity of $0.8 \times 10^{12}$ protons.

In the first 100 μs after injection,

Fig. 3(a) the SPD signal and PSDRV *vs.* time.

Fig. 3(b) the SPD signal and RPOS *vs.* time.

In the first 1.6 ms after injection,

Fig. 3(c) the SPD signal and PSDRV *vs.* time.

Fig. 3(d) the SPD signal and RPOS *vs.* time.

Fig. 3(e) the SPD signal and FPERR *vs.* time.

Fig. 3(f) the SPD signal and charge *vs.* time.

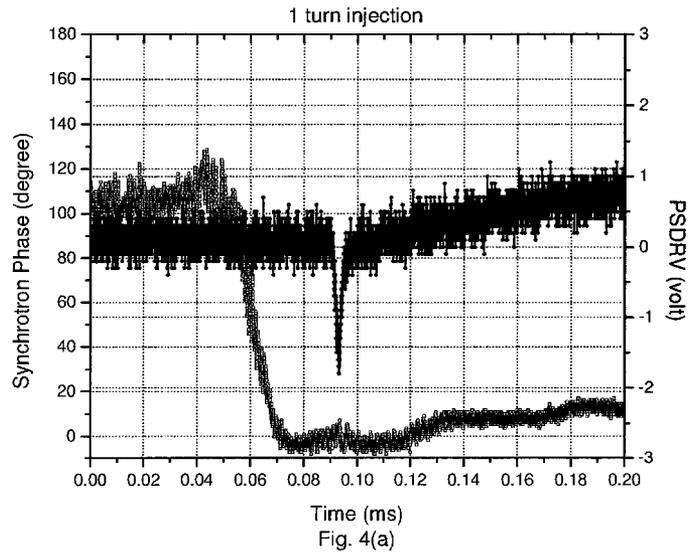

Fig. 4(a)

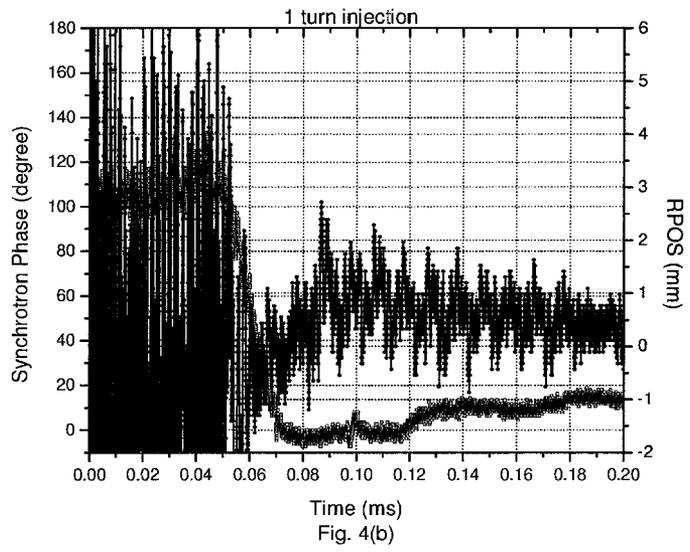

Fig. 4(b)

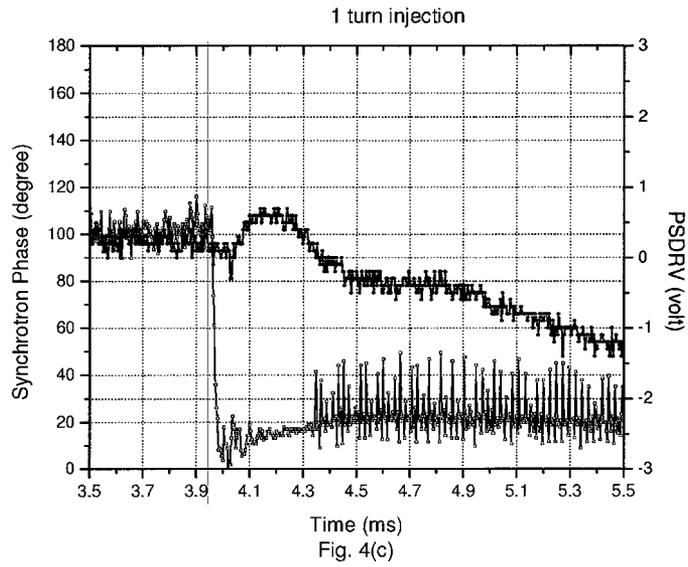

Fig. 4(c)

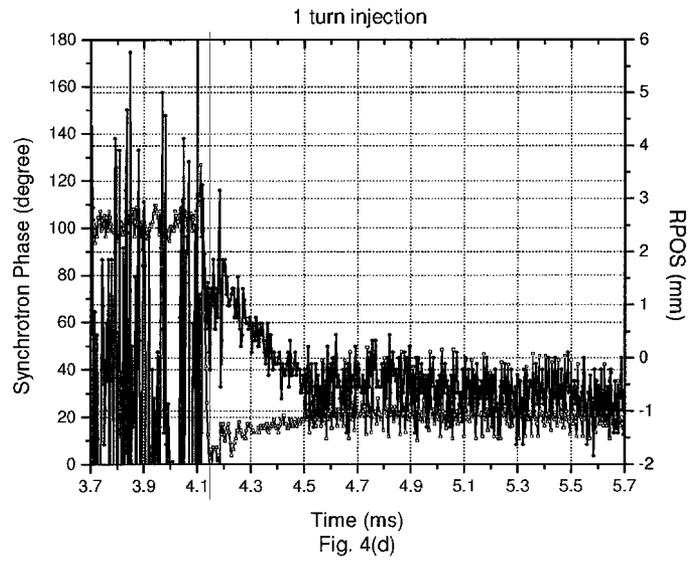

Fig. 4(d)

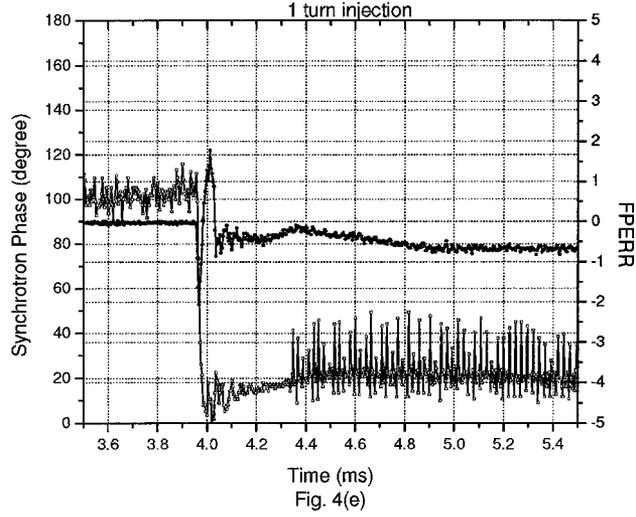

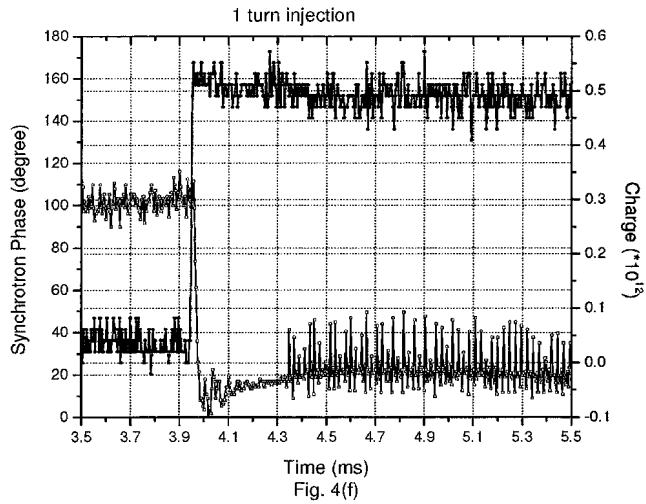

Fig. 4 The data were taken in the matched situation at injection for the extracted beam intensity of $0.4 \times 10^{12}$ protons.

In the first 100 μs after injection,

Fig. 4(a) the SPD signal and PSDRV *vs.* time.

Fig. 4(b) the SPD signal and RPOS *vs.* time.

In the first 1.6 ms after injection,

Fig. 4(c) the SPD signal and PSDRV *vs.* time.

Fig. 4(d) the SPD signal and RPOS *vs.* time.

Fig. 4(e) the SPD signal and FPERR *vs.* time.

Fig. 4(f) the SPD signal and charge *vs.* time.